\documentstyle[preprint,tighten,eqsecnum,floats,aps]{revtex}

\def\C{{\cal C}}
\def\B{{\cal B}}
\def\E{{\cal E}}
\def\I{{\cal I}}
\def\G{{\cal G}}
\def\Q{{\cal Q}}
\def\M{\hat{M}}
\def\g{\hat{g}}
\def\h{\hat{h}}
\def\K{\hat{K}}
\def\R{\hat{R}}
\def\T{\hat{T}}
\def\t{\hat{\tau}}
\def\S{\hat{S}}
\def\L{\Lambda}
\def\l{\ell}
\def\O{\Omega}
\def\D{\nabla}
\def\Gn{G_{(d)}}
\def\bar{\overline}

\def\ba{\begin{eqnarray}}
\def\ea{\end{eqnarray}}
\def\be{\begin{equation}}
\def\ee{\end{equation}}
\def\={\mathrel{\widehat\mathalpha{=}}}
\def\ads{anti-de Sitter }

\preprint{\vbox{\baselineskip=12pt
\rightline{CGPG-99/11-8}}}

\begin{document}
\draft
\title{Asymptotically Anti-de Sitter Space-times:\\
Conserved Quantities}
\author {Abhay\ Ashtekar and Saurya\ Das}
\address{Center for Gravitational Physics and Geometry \\
Department of Physics, The Pennsylvania State University \\
University Park, PA 16802-6300, USA}
\maketitle

\begin{abstract}

Asymptotically \ads space-times are considered in a general dimension
$d\ge 4$. As one might expect, the boundary conditions at infinity
ensure that the asymptotic symmetry group is the \ads group (although
there is an interesting subtlety if $d=4)$.  Asymptotic field
equations imply that, associated with each generator $\xi$ of this
group, there is a quantity $Q_\xi$ which satisfies the expected
`balance equation' if there is flux of physical matter fields across
the boundary $\I$ at infinity and is absolutely conserved in absence
of this flux.  Irrespective of the dimension $d$, all these quantities
vanish if the space-time under considerations is (globally) anti-de
Sitter.  Furthermore, this result is required by a general covariance
argument.  However, it contradicts some of the recent findings based
on the conjectured ADS/CFT duality. This and other features of our
analysis suggest that, if a consistent dictionary between gravity and
conformal field theories does exist in fully non-perturbative regimes,
it would have to be more subtle than the one used currently.

\pacs{Pacs: 04.20.-q, 04.20.Ha, 11.25.-w}
\end{abstract}

\section{Introduction}
\label{sec1}

In the early to mid eighties, interest in \textit{asymptotically} \ads
space-times was sparked by the discovery that \ads space-times
generically arise as ground states in certain supergravity theories
which at the time were considered to be among the most promising
candidates for quantum gravity.  In particular, the relevant boundary
conditions at infinity, asymptotic symmetries and associated conserved
quantities were discussed by a number of authors.  (See, e.g,.
\cite{ad,bf,h,ghhp,bgh,am,ht}).  Some of the early investigations used
intuitively natural but somewhat imprecise boundary conditions based
on the behavior of metric coefficients in certain coordinate systems.
These coordinates referred (sometimes implicitly) to an underlying
\ads metric to which the given physical metric was to approach.
Unfortunately, given a candidate asymptotically \ads space-time, there
are ambiguities in picking out `the' \ads metric to which the given
metric approaches.  As a result, the definitions of conserved
quantities had certain ambiguities.%
\footnote{There is an analogous problem at null infinity in the
asymptotically Minkowskian context.  In presence of generic
gravitational radiation, it is not possible to single out `the'
Minkowski metric to which a physical metric approaches and this leads
to the well-known super-translation ambiguities in the definition of
angular-momentum at null infinity \cite{w}.  However, that
problem is `intrinsic' in the sense that it is tied to the enlargement
of the asymptotic symmetry group from the Poincar\'e group to the
Bondi-Metzner-Sachs group \cite{bms}.  In the asymptotically \ads
context, all treatments (correctly) assumed that the asymptotic
symmetry group was the \ads group and the ambiguities arose simply
because of imprecision in techniques; limits to infinity are now
delicate because some of the metric coefficients diverge there.}

In \cite{am}, it was pointed out that these problems can be avoided by
using Penrose's conformal techniques \cite{p}.  If, in addition to the
restrictions required in Penrose's conformal completion, one imposes
`reflective boundary conditions', the boundary $\I$ becomes
conformally flat and the asymptotic symmetry group reduces to the
group of its conformal isometries, i.e., the \ads group $O(d-1, 2)$ in
four dimensions\cite{h,am}.  A detailed analysis of the asymptotic
field equations showed that it is possible to define conserved
quantities as 2-sphere integrals involving the electric part of the
Weyl tensor and conformal Killing fields on $\I$ representing
infinitesimal asymptotic symmetries.  In particular, since the Weyl
tensor vanishes identically in the \ads space-time, all conserved
quantities vanish as well.  Finally, there exists a rigorous analysis
of the mixed initial-value boundary-value problem within Einstein's
theory which shows that there exists a large class of asymptotically
\ads solutions satisfying these boundary conditions \cite{f}.

The investigations mentioned above were carried out in four space-time
dimensions.  During the last three years, higher dimensional
asymptotically \ads space-times have attracted a great deal of
attention because of a bold conjecture put forward by Maldecena
\cite{m}.  In particular, inspired by the anti-de Sitter/conformal
field theory (ADS/CFT) correspondence proposed in this conjecture, the
issue of conserved quantities in such space-times was investigated in
higher dimensions \cite{old,bk,new}.  The starting point of these
analyses is the introduction of an appropriate action.  As in
asymptotically Minkowski space-times, the Einstein-Hilbert action of
asymptotically \ads space-times diverges \cite {hh} (even when one
restricts oneself to a space-time region bounded by two space-like
surfaces with a finite time-like separation).  The strategy
\cite{bk,new} is to use a `counter-term subtraction method', where the
counter-terms depend on the geometry of the space-time boundary, which
in turn encodes the property that the space-time under considerations
is asymptotically anti-de Sitter.  By varying the action with respect
to the boundary metric, one obtains an `effective stress-energy
tensor' on the boundary.  Note that, contrary to the normal usage of
the term `stress-energy' in the general relativity literature, this
tensor field does not refer to any physical matter fields near the
boundary, but is meant to encode the purely gravitational contribution
to the conserved quantities.  More precisely, the conserved quantities
are obtained by first computing $(d-2)$-sphere integrals of this
effective stress energy tensor, contracted with vector fields
representing asymptotic symmetries, and then taking the limit, as the
boundary goes to infinity.

These methods have led to some intriguing results on the relation
between conserved quantities calculated in general relativity and
analogous quantities calculated using an appropriate conformal field
theory on the boundary.  In particular, while the energy of the pure
\ads space-time does vanish when $d=4$ as in earlier analyses
\cite{am}, it fails to vanish for $d=5$.  Thus, using these methods,
\textit{from purely gravitational considerations} one finds that there
is a `vacuum energy' in 5-dimensional asymptotically \ads space-times.
This seems puzzling from the conventional general relativity
standpoint. However, it has been argued that the ADS/CFT
correspondence illuminates the origin of this vacuum energy: its value
agrees \textit{precisely} with the Casimir energy in the dual (${\cal
N}=4$, super-symmetric) Yang-Mills theory on the boundary used in the
Maldecena conjecture.  This agreement is striking because the two
calculations are so different.  The calculation on the gravity side is
non-perturbative; it refers to the full non-linear theory rather than
the linearized theory. The Casimir energy on the Yang-Mills side, on
the other hand, uses only linear quantum fields.  Even more surprising
is the feature that the first calculation is purely classical while
the second refers to a quintessential quantum effect.  In particular,
the value of the Casimir energy is proportional to $\hbar$ and has no
analog in the classical Yang-Mills theory.  And yet, the available
dictionary in the conjectured ADS/CFT correspondence converts the
relevant parameters in such a way that the quantum Casimir energy of
gauge fields is translated \textit{precisely} to the classical vacuum
energy in general relativity.  Therefore, the agreement has been used
as a a clear example of the power and subtlety of the ADS/CFT duality
which interpolates between a gravity theory in the bulk and a
Yang-Mills theory on the boundary.

It is therefore desirable to understand various aspects of this
calculation as thoroughly and as deeply as possible.  The purpose of
this note is to re-examine the gravity side of the calculation using
Penrose's conformal methods which, as discussed above, avoid a number
of pitfalls associated with boundary conditions and limiting
procedures.

In Sections \ref{s2} and \ref{s3}, we extend the analysis of \cite{am}
to higher dimensions.  We find that all results of \cite{am} directly
generalize to $d>4$ (although there is a small but interesting
difference in in the treatment of asymptotic symmetries). In
particular, in any dimension $d\ge 4$, we find that all conserved
quantities at $\I$ vanish identically in pure \ads space-times.  At
first sight, whether these quantities vanish or not may seem to depend
on how one chooses their `ground state values' which may in turn
depend on one's approach to the problem.  However, one can give a
simple argument to show that, if the definition of these quantities is
covariant, they must vanish in the \ads space-time, irrespective of
the method used to define them.  At its crux, it is the same reasoning
that one invokes to conclude that the angular momentum of a
spherically symmetric configuration must vanish.

There is thus a tension between our results and those inspired by the
ADS/CFT correspondence.  To clarify the issue, in Section \ref{s4} we
explicitly calculate the difference between our conserved quantities
and those constructed using the `counter-term subtraction procedure'.
Although the starting points are quite different, we show that, if
$d=4$, the two definitions agree for \textit{general} asymptotically
\ads space-times (i.e., not just for Kerr-\ads space-times for which
explicit calculations are available).  If $d=5$, as one would expect,
the two definitions disagree but the difference is finite in
\textit{all} asymptotically \ads space-times.  We provide an explicit
expression for the difference.  In the Schwarzschild-\ads family, the
difference is a constant shift by the vacuum energy.  However, as far
as we can tell, for more general asymptotically \ads space-times,
there are some additional terms. These differences suggest that the
ADS/CFT dictionaries currently used to go between gravity theories in
the bulk and conformal field theories on the boundary have certain
unsatisfactory features in the non-perturbative regimes. This issue is
discussed in Section \ref{s5}.

This note is addressed to both general relativity and string theory
communities.  In the hope of to bridging some apparent gaps, we have
included introductory remarks on the ADS/CFT conjecture and some
details on higher dimensional gravity and conformal techniques.

\section{Asymptotically anti-de Sitter space-times}
\label{s2}

This section is divided in to two parts.  In the first, we present the
basic definition of asymptotically \ads space-times and illustrate the
Penrose completion with a simple example.  In the second, we work out
the basic consequences of the conditions introduced in the definition.
Throughout this section, we work in d-dimensional space-times with
$d\ge 4$.  (The $d=3$ case needs a special treatment already in the
asymptotically Minkowskian context \cite{abs} and will be discussed
elsewhere.)  For simplicity, all fields will be assumed to be smooth
(i.e., $C^\infty$) on the domain of their definitions.

\subsection{Definition}
\label{s2.1}

A $d$-dimensional space-time $(\M,\g_{ab})$ will be said to be
\textit{asymptotically \ads} if there exists a manifold $M$ with
boundary $\I$, equipped with a metric $g_{ab}$ and a diffeomorphism
from $\M$ onto $M - \I$ of $M$ (with which we identify $\M$ and $M
-\I$) and the interior of $M$ such that:
\begin{enumerate}
\item there exists a function $\O$ on $M$ for which $g_{ab} = \O^2 
\g_{ab}$ on $\M$;

\item $\I$ is topologically $S^{d-2}\times R$, $\O$ vanishes on $\I$
but its gradient $\D_a \O$ is nowhere vanishing on $\I$;

\item On $\M$, $\g_{ab}$ satisfies $\R_{ab} -\frac{1}{2}\R\, \g_{ab}
+\L \g_{ab} = 8\pi \Gn \T_{ab}$, where $\L$ is a negative constant,
$\Gn$ is Newton's constant in $d$-dimensions, and the matter
stress-energy $\T_{ab}$ is such that $\O^{2-d} \T_{ab}$ admits a
smooth limit to $\I$.

\item The Weyl tensor of $g_{ab}$ is such that $\O^{4-d}C_{abcd}$ is
smooth on $M$ and vanishes at $\I$.

\end{enumerate}

The meaning of the conditions in the definition is as follows.  The
first condition ensures that the physical metric $\g_{ab}$ is
conformally related to the new metric $g_{ab}$ which is mathematically
more convenient.  The second condition ensures that the topology of
the boundary is the one suggested by the geometry of \ads space-times.
Since the conformal factor $\O$ vanishes on $\I$, the boundary
attached in the conformal completion is at infinity with respect to
the physical metric $\g_{ab}$.  The condition that $\D_a\O$ is nowhere
vanishing on $\I$ ensures that $\O$ can be used as a `good (radial)
coordinate' in a neighborhood of $\I$ in the completed space-time
$(M,g_{ab})$.  (In terms of the physical metric, heuristically, the
condition tells us that $\O$ `falls off as $1/r$'.)  The conformal
rescaling brings the boundary (at infinity of $(\M,\g_{ab})$) to a
finite `distance' (with respect to $g_{ab}$).  Therefore, the delicate
operations involving limits of various fields in $\M$ can now be
reduced to simple \textit{local} differential geometry of fields at
$\I$ in the completed space-time.  This simple fact removes
ambiguities in the limiting procedures and streamlines manipulations
at infinity.  

The third condition restricts the asymptotic behavior of matter
fields. (In the analyses based on the `counter-term subtraction' it is
implicitly assumed that there is no physical matter in a
\textit{neighborhood} of infinity, whence this condition is trivially
met.)  The fall-off condition on $\T_{ab}$ is motivated by the
analysis of test fields in \ads space-time; it ensures that the fluxes
of conserved quantities associated with test fields across $\I$ are
well-defined.  We will see that for the coupled Einstein-matter field
system, it will also ensure that we obtain the physically expected
`balance equations' for conserved quantities when there is flux of
matter fields across $\I$. Finally, since the Weyl tensor vanishes
identically in the anti de-Sitter space-time, it is natural to expect
it to fall off at an appropriate rate in asymptotically \ads
space-times. The precise rate is given in the fourth condition. It is
suggested by the following dimensional considerations. One expects the
leading non-zero contribution to the components of the Weyl tensor in
an orthonormal tetrad to behave asymptotically as $\Gn M/ r^n$ for
some $n$, where $M$ is a measure of the total mass. Physical
dimensions of the Weyl tensor and $\Gn$ dictate $n = d-1$. Using the
behavior $\O \sim 1/r$ and the conformal rescalings of orthonormal
tetrads in the passage from ${\hat g}_{ab}$ to $g_{ab}$ we arrive at
the fall off given in the fourth condition. However, this condition
can be weakened because the first three conditions already constrain
the asymptotic Weyl curvature. These and other details will be
discussed elsewhere.

Next, let us consider a simple example to illustrate how this
definition works.  Let $(\M,\g_{ab})$ be Schwarzschild-\ads
space-time.  Then, in an asymptotic region (away from the horizon) the
metric $\g_{ab}$ can be taken to be
\ba\label{2.1}
\g_{ab} dx^a dx^b &=& - \hat{F}^2(r) dt^2 + \hat{F}^{-2}(r) dr^2
+ r^2 d\sigma^2\, ,
\nonumber\\
{\rm with} \quad \hat{F}^2 (r) &=& 1 + \frac{r^2}{\l^2} - 
(\frac{r_o}{r})^{d-3}
\ea
where $d\sigma^2$ is the line-element of the round, unit
$d-2$-sphere metric. Following the usual conventions we have set
\be \label{2.2}
\L = - \frac{(d-1)(d-2)}{2 \l^2} \quad {\rm and} \quad
M_{(d)} = \frac{(d-2)(d-3) A_{(d-2)} r_0^{d-3}}{16\pi G_d}
\ee
where $M_{(d)}$ is the Schwarzschild mass and $A_{(d-2)}$ the area of
the unit $(d-2)$-sphere.  To show that this space-time satisfies our
boundary conditions, we need to first pick a conformal factor $\O$ and
then carry out the completion by attaching to $\M$ the $\O =0$
surface. Set $\O = 1/r$.  By replacing $r$ by $1/\O$ in (\ref{2.1})
we obtain
\ba
g_{ab}dx^a dx^b &:=& \O^2\g_{ab}dx^a dx^b 
= -{F}^2(\O) dt^2 + F^{-2} (\O) d\O^2 + d\sigma^2\, , 
\nonumber \\
{\rm with}\quad F^2(\O) &=& \O^2 \hat{F}^2(r) = 
(\frac{1}{\l^2} + \O^2 - r_o^{d-3}\O^{d-1} ) 
\ea
The surface $\O =0$ did not belong to $\M$ since it corresponds to
$r=\infty$; indeed, if we simply attached it to $\M$, the metric
$\g_{ab}$ would be singular there.  However, in the $(t,\O, {\rm
angles})$ chart, it is trivial to extend $\M$ by attaching to it a
boundary $\I$ defined by $\O =0$. The rescaled metric $g_{ab}$ is
well-defined at this boundary.

Let us now set $M = \M \cup \I$ and verify that the conditions of the
definition are satisfied.  Clearly, $g_{ab}$ is smooth on $M$ (i.e.,
its components in the chart $(t,\Omega, {\rm angles})$ are all
smooth).  $\I$ is topologically $S^{d-2}\times R$; the conformal
factor $\Omega$ vanishes at $\I$ but its gradient $\D_a\Omega$ does
not (indeed, $g^{ab} \D_a\O \D_b\O = {1}/{\ell^2}$ on $\I$).  The
third condition is trivially satisfied since the stress energy tensor
$\T_{ab}$ of physical matter fields vanishes identically on $\M$.
Finally, a direct calculation or the dimensional argument given above
shows that the condition on Weyl curvature is also satisfied.

We will conclude with some remarks on the conformal freedom. First,
note that we could choose the conformal factor $\O$ to be ${1}/{r}$ in
\textit{all} dimensions; there is no $d$-dependence in the power of
$r$. However, there is considerable freedom in the choice of $\O$. In
particular, $\bar\O = \omega \O$ is an equally good conformal factor
provided $\omega$ is a smooth, nowhere vanishing function on $M$. In
Schwarzschild-\ads space-times, the choice $\O = {1}/{r}$ is adapted
to the isometries and, in particular, to the rest frame selected by
the time translation Killing field $\partial/\partial t$. In the \ads
space-time, on the other hand, there is no natural rest frame and
choices of conformal factors associated with distinct frames are
related by a $\omega$ which is non-vanishing on $\I$. Because of this
conformal freedom, for physical quantities defined at $\I$, covariance
with respect to the physical space-time is tied to conformally
invariance at $\I$. This point will play an important role in Section
\ref{s3}.

\subsection{Asymptotic fields and their equations}
\label{s2.2}

We now wish to analyze physical fields near $\I$ and explore equations
they satisfy. These equations will lead us to the asymptotic symmetry
group and conserved quantities in Section \ref{s3}.

Let us begin by introducing some notation. The Riemann tensor
of the metric $g_{ab}$ can be decomposed in to the Weyl and Ricci 
tensors as follows:
\be\label{2.6}
R_{abmn} = C_{abmn} + \frac{2}{d-2} \left( g_{a[m}S_{n]b} -
g_{b[m}S_{n]a}\right)\, ,
\ee
where $S_{ab}$ is given by
\be\label{2.7}
S_{ab} = R_{ab} - \frac{R}{2(d-1)} g_{ab}\, . 
\ee
Next, under the conformal rescaling $g_{ab} = \O^2 \g_{ab}$ on $\M$,
the Weyl tensor transforms trivially, $\hat{C}_{abc}{}^d =
C_{abc}{}^d$, while the Ricci pieces transform via
\be\label{2.8}
{\hat{S}_{ab}} = S_{ab} + (d-2) \Omega^{-1} \nabla_a n_b - 
\left(\frac{d-2}{2}\right) \Omega^{-2} g_{ab} n^c n_c \, ,
\quad{\rm where}\quad  n_a := \nabla_a \O 
\ee
Field equations enable us to express
$\S_{ab}$ in terms of the cosmological constant and the matter
stress-energy as
\be\label{2.9}
{\hat{S}_{ab}} - \frac{\Lambda}{d-1} \g_{ab} = 
8\pi \Gn \left({\hat T}_{ab} -
\frac{{\hat T}}{d-1} \right) {\g}_{ab} 
\ee

Multiplying (\ref{2.8}) by $\O^2$, taking the limit as $\O$ tends
to $0$ and  using the fact that $\O^{2-d}\T_{ab}$ admits a smooth 
limit to $\I$, we obtain:
\be\label{2.10}
n \cdot n \= \frac{-2 \Lambda}{(d-2)(d-1)} \equiv \frac{1}{\l^2}\, , 
\ee
where $\=$ denotes equality restricted to $\I$.  (We will use this notation 
extensively in the rest of the paper.)  Thus, $n^a$ is space-like at $\I$, 
whence $\I$ is a time-like surface.  Now, if a smooth field $f$ vanishes on 
$\I$, then ${\bf f} = \O^{-1} f$ admits a smooth limit to $\I$.  Hence it 
follows that $\O^{-1}\, (n\cdot n - 1/\l^2)$ admits a smooth limit.  Now, 
multiplying (\ref{2.9}) by $\O$ and taking limit to $\I$ we obtain
\be\label{2.11} 
2 \nabla_a n_b \= \left[\lim_{{ } \rightarrow
\I}~{\Omega^{-1} \left(n^c n_c - \frac{1}{l^2} \right)}\right] g_{ab}
\= \frac{2}{d} (\nabla_c n^c) g_{ab} 
\ee
Next, it is easy to verify that under a conformal rescaling $\bar\O
= \omega \O$, we have
\be\label{2.12} 
{\bar \nabla}_a {\bar n}_b \= \frac{1}{d}\,\omega
(\nabla_c n^c) g_{ab} + (n^c \nabla_c \omega) g_{ab} \ee
where $\bar\D$ is the derivative operator compatible with
$\bar{g}_{ab} = \bar{\O}^2 \hat{g}_{ab}$ and $\bar{n}_a = \D_a
\bar{\O}$. Hence, by suitable rescaling, we can \textit{always} obtain
a conformal completion in which
\be\label{2.13}
\nabla_a n_b \= 0
\ee
\textit{From now on, we will only use such conformal completions.}
(The explicit completions of the Schwarzschild-\ads space-times given
in Section \ref{s2.1} meet this condition.)  There is still a
remaining conformal freedom: $\O \mapsto \omega \O$ where
$n^a\D_a\omega \= 0$. \textit{Note that the value of $\omega$ at $\I$
is unrestricted.}

The first key equation for obtaining conserved quantities results from
taking the derivative of (\ref{2.8}) and using Einstein's equation
(\ref{2.9}):
\be\label{2.14}
\Omega \nabla_{[a} S_{b]c} + \left(\frac{d-2}{2}\right)
 C_{abc}^{~~~d}~n_d = 
8\pi \Gn \tilde{T}_{d[a} g_{b]c}~n^d + 
8\pi \Gn \D_{[a}\O\, \tilde{T}_{b]c} 
\ee
where $\tilde{T}_{ab} = \T_{ab} - ({\T}/{d-1}) g_{ab}$. The second
key equation is the (contracted) Bianchi identity on $(M, g_{ab})$
\be \label{2.17}
\D^d C_{abcd} + \frac{2(d-3)}{(d-2)}\,\, \D_{[a}S_{b]c} =0\, .
\ee
We can now eliminate the $S_{ab}$ term in (\ref{2.14}) using
(\ref{2.17}). Since the fourth condition in the main definition
ensures that
\be \label{2.18} 
{\bf K}_{abcd} := \lim_{{ } \rightarrow\I}
\O^{3-d}C_{abcd} 
\ee
admits a smooth limit to $\I$, using the fall-off condition on
$\T_{ab}$ given in the main definition, we obtain
\be\label{2.19}
\D^d K_{abcd} \= \lim_{{ } \rightarrow\I} - \frac{2(d-3)}{(d-2)}\,\,
\O^{2-d}\,\, 8\pi \Gn \, \left[ \tilde{T}_{d[a} g_{b]c} n^d
+ \D_{[a} (\O \tilde{T}_{b]c}) \right]
\ee
Finally, by projecting this equation via $n^an^c h^b{}_m$, where
$h_{ab} \= g_{ab} -\l^2 n_an_b$ is the intrinsic metric on $\I$, we
obtain,
\be\label{2.20}
D^d \E_{md} =\, - {8\pi \Gn}\,\, (d-3)\,\, 
{\bf T}_{ab} n^a h^b{}_m \,.
\ee
Here, $D$ is the intrinsic derivative operator on $\I$ compatible with
$h_{ab}$, $\E_{ab} \= \ell^2 {\bf K}_{ambn}n^m n^n$ is the electric
part of the (leading order) Weyl tensor ${\bf K}_{amnb}$ at $\I$ and
${\bf T}_{ab} = \lim \O^{2-d} \T_{ab}$. (Note that ${\E}_{ab}$ and
${\bf T}_{ab}$ are smooth fields at $\I$.) This is the key identity
that will enable us to introduce conserved quantities at infinity.

\section{Symmetries and associated charges}
\label{s3}

This section is divided in to two parts. In the first, we discuss
asymptotic symmetries and in the second we introduce the conserved
quantities associated with these symmetries.

\subsection{Asymptotic symmetries}
\label{s3.1}

In the physical space-time picture, the asymptotic symmetry group $\G$
is the quotient $({\rm Diff}\, \M /{\rm Diff}_o \M)$ of the group of
diffeomorphism which preserve the boundary conditions by its sub-group
consisting of asymptotically identity diffeomorphisms.  In the
conformally completed space-time, $\G$ is the subgroup of the
diffeomorphism group on $\I$ which preserves the `universal structure'
at $\I$, i.e., the structure that is common to boundaries of
\textit{all} asymptotically \ads space-times. Let us begin by
exploring this structure.

Since $\D_an_b \= 0$, it follows that the extrinsic curvature $K_{ab}$
of $\I$ vanishes. Hence, the intrinsic curvature tensor $\R_{abcd}$ of
$(\I, h_{ab})$ is given by: $\R_{abcd} \= h_a{}^m h_b{}^n h_c{}^s
h_d{}^t R_{mnst}$. Now, our boundary conditions imply that the Weyl
tensor $C_{abcd}$ of $g_{ab}$ vanishes on $\I$ for all $d\ge
4$. Hence, using (\ref{2.6}) it now follows that the
\textit{intrinsic} Weyl tensor $\C_{abcd}$ of $\I$ vanishes. Thus,
\textit{if} $d>4$, $(\I, h_{ab})$ is conformally flat in \textit{all}
asymptotically \ads space-times. Hence $\G$ is just the $d(d+1)/2$
dimensional conformal group, i.e., the \ads group in $d$ dimensions.

If $d=4$, however, $\I$ is a 3-manifold and vanishing of $\C_{abcd}$
imposes no restriction at all.  Let us examine this case in some
detail. Note first that, in any dimension $d>4$, we used only the
first three conditions in the main definition to derive all equations
up to (\ref{2.17}). Therefore, using (\ref{2.14}) and the fall-off
conditions on the stress energy $\T_{ab}$, we can conclude
$C_{abcd}n^d \=0$ without any reference to the fourth condition in the
definition on the fall-off of Weyl tensor. This in particular implies
that the electric and magnetic parts of the Weyl tensor $C_{abcd}$
must vanish at $\I$. Now, in four dimensions, the Weyl tensor has no
other independent components and hence we can conclude
$C_{abcd}\=0$. Therefore, if $d=4$, the fourth condition in the
definition is redundant. However, as noted above, in this case, the
definition does \textit{not} imply that the intrinsic geometry of $\I$
is conformally flat. In the \ads space-time, on the other hand, this
geometry is indeed conformally flat.

Therefore, to fully capture the idea that $(\M, \g_{ab})$ is
asymptotically \ads in four dimensions, we need to impose a new
condition which may be regarded as a replacement of the fourth
condition in the definition.  Let us suppose that \textit{the magnetic
part}
\be
\B_{ab} :={}^\star\! {\bf K}_{ambn}n^mn^n \equiv \lim_{~\mapsto \I}
\O^{-1} {}^\star C_{ambn}n^mn^n 
\ee
\textit{of the asymptotic Weyl curvature} ${\bf K}_{ambn}$
\textit{vanishes on} $\I$.  Then, one can show that $(\I, h_{ab})$ is
conformally flat and $G$ is the \ads group in four dimensions
\cite{am}.  (A completely analogous condition is necessary to obtain
Poincar\'e group as the asymptotic symmetry group at spatial infinity
of asymptotically Minkowskian space-times \cite{ah,ar}.)  The
condition $\B_{ab} \=0$ is sometimes referred to as the `reflective
boundary condition' \cite{h}.

Thus, as one might intuitively expect, the asymptotic symmetry group
is the \ads group in all cases (i.e., for all $d\ge 4$).  Note that,
since $\I$ is not endowed with additional universal structure, $\G$
can not be further reduced.  In any one completion, of course, one can
single out the isometry group of $h_{ab}$ as a sub-group of $\G$.
However, this group, and indeed even its dimension, will vary with the
choice of the conformal factor (which determines $h_{ab}$).  It is
only the equivalence class $[h_{ab}]$ of conformally flat metrics on
$\I$ and the full conformal isometry group $\G$ that have an invariant
meaning.  On the other hand, if one restricts oneself to a specific
space-time $(\M,\g_{ab})$ admitting isometries, in any conformal
completion of \textit{that} space-time, one can select a preferred
subgroup of $\G$ (since every Killing vector of $\g_{ab}$ admits an
extension to $\I$ and is a conformal Killing field of $h_{ab}$
irrespective of the choice of $\Omega$.)

\subsection{Conserved quantities}
\label{s3.2}

Eq (\ref{2.20}) is a differential conservation law which immediately
leads us to the required `conserved' charges. Given any infinitesimal
asymptotic symmetry (i.e., a conformal Killing field $\xi^a$ on $\I$)
and a $d-2$ sphere cross section $C$ on $\I$ we can now define a
`conserved' quantity
\be \label{3.1}
Q_\xi [C] := - \frac{1}{8\pi \Gn}~\frac{\l}{n-3}~~\oint_C 
{\cal E}_{ab} \xi^a dS^b \ee
It now follows immediately from (2.20) that $Q_\xi$ satisfies a
balance equation: Given two cross-sections $C_1$ and $C_2$ which form
boundaries of a $d-1$ dimensional region $\Delta\I$ of $\I$, we have
\be \label{3.2}
Q_\xi [C_2] - Q_\xi [C_1] = \int_{\Delta \I} {\bf T}_{ab} \xi^a dS^b
\ee
where we have used the fact that ${\cal E}_{ab}$ is trace-free. Thus
the difference between values of $Q_\xi$ evaluated at the two
cross-sections is precisely the flux of that physical quantity,
associated with matter, across the portion $\Delta\I$ of $\I$. It is
only when the matter flux vanishes that the quantity is absolutely
conserved (whence the use of inverted commas in the label
`conserved').

As in Section \ref{s2.1}, let us consider the example of a
Schwarzschild-anti-de Sitter space-time. Since there is no physical
matter field anywhere, all the $Q_\xi$ are absolutely conserved. It
is straightforward to evaluate them explicitly. If the conformal
Killing field $\xi$ corresponds to the time-translation Killing field
on $({\hat M}, {\hat g}_{ab})$, then $Q_\xi=M$, the mass parameter in
the metric. All other conserved quantities vanish. In particular, if
$({\hat M}, {\hat g}_{ab})$ is the anti-de Sitter space-time, all
$Q_\xi$ vanish identically, irrespective of the dimension $d$.

We will now present a general argument based on covariance to show
that this last conclusion is robust and not tied to the specific
procedure we used to define conserved quantities.  Let us suppose that
one has a covariant procedure that leads to a conserved quantity
${\tilde Q}_\xi$ associated with each conformal Killing field $\xi^a$
on $\I$ (i.e., a procedure which uses only the boundary conditions and
does not refer to additional structures such as preferred coordinates
or background fields.)  Let us first fix a general asymptotically
anti-de Sitter space-time $({\hat M}, {\hat g_{ab}})$ in which there
are no matter fields near infinity and compute its $d(d+1)/2$
absolutely conserved charges ${\tilde Q}_\xi$.  Apply a diffeomorphism
$\varphi$ which is an asymptotic symmetry, i.e., defines an element of
$\G$.  The image ${\hat g}^\prime_{ab} = \varphi^*({\hat g}_{ab})$ of
${\hat g}_{ab}$ is again asymptotically anti-de Sitter and, by
covariance, values $\tilde{Q}^\prime$ of its conserved charges are
$\tilde{Q}^\prime_\xi = \tilde{Q}_{\varphi \cdot \xi}$ where $\varphi
\cdot \xi$ is given by the action of $\G$ on $\I$.  (More precisely,
the $d(d+1)/2$ charges $\tilde{Q}_\xi$ define an `anti-de Sitter
momentum' which lies in the dual of the Lie algebra of $\G$ and the
anti-de Sitter momenta of ${\hat g}_{ab}$ and ${\hat g}^\prime_{ab}$
are related by the natural action of $\G$ on the dual of its Lie
algebra.)  Now, let $({\hat M}, {\hat g}_{ab})$ be the anti-de Sitter
space-time itself and choose for $\varphi$ an {\it isometry} of ${\hat
g}_{ab}$.  Then ${\hat g}^\prime_{ab} = {\hat g}_{ab}$.  Hence we must
have $Q^\prime_{\varphi \cdot \xi} = Q^\prime_{\xi}$ for all $\varphi$
in $\G$.  This is possible only if $Q_\xi=0$.%
\footnote{Note that (like the rest of our analysis) this argument does
not go through if $d=3$. In this case, there are two possible sets of
boundary conditions at infinity \cite{bh} and neither yields $O(2,2)$
as the asymptotic symmetry group. With the more commonly used weaker
set, the asymptotic symmetry group is \textit{infinite} dimensional
and does not admit $O(2,2)$ as an invariant sub-group. Therefore, the
last line in our argument need not hold. This issue will be discussed
in detail elsewhere.}

The crux of the argument is quite simple. In the asymptotically
Minkowskian context, the same argument tells us that if the total
$4$-momentum at spatial infinity is defined by a covariant procedure,
then it must vanish in Minkowski space since the Minkowski metric is
invariant under Lorentz transformations and there is no non-zero
Lorentz invariant $4$-vector.

We will conclude this section with three remarks.\\ 
1.  The anti-de Sitter space-time is conformally flat. Therefore, it
was natural to require the Weyl curvature of {\it asymptotically}
anti-de Sitter space-times to vanish at infinity. The physical
observables $Q_\xi$ are defined using its `residue' near infinity.  By
contrast, the extraction of this physical information from metric
components is much more difficult and delicate: since the components
of the anti-de Sitter metric diverge at infinity in standard charts,
the extraction involves a comparison between two infinite
quantities. Note also that even in the asymptotically Minkowskian
context, one can define energy-momentum and angular-momentum at
spatial infinity in terms of the asymptotic behavior of the Weyl
tensor \cite{ah,ar}.  Furthermore, this approach offers perhaps the
clearest way to weed out the unwanted super-translations.\\
 2.  If there are no physical matter fields near $\I$, all $d(d+1)/2$
quantities $Q_\xi$ are absolutely conserved.  This is in striking
contrast with the situation at null infinity of asymptotically
Minkowskian space-times, where, even in the absence of matter,
gravitational radiation can and does carry away energy-momentum and
angular momentum (see, e.g., \cite{bms,p,wald}).  On the other hand,
matter fields can carry away these quantities in both cases.  Thus, in
certain respects the boundary $\I$ in the asymptotically \ads context
is analogous to spatial infinity \cite{ah,ar} in the asymptotically
Minkowskian space-times and in other respects it is analogous to null
infinity \cite{bms,p,wald}.\\
3. Finally, note that since $\xi^a$ are \textit{conformal} Killing
fields on $(\I, h_{ab})$, $Q_\xi$ would not be absolutely conserved
even in absence of matter if $\E_{ab}$ had not been trace-free.
Thus, in that case, conserved charges would not exist. This is quite
distinct from the issue of whether or not the Poisson algebra between
conserved charges yields a `normal' or an `anomalous' representation
of the Lie algebra of the asymptotic symmetry group.

\section{Comparison}
\label{s4}

In this section we will present an explicit comparison between the
conserved charges $Q_\xi$ of Section \ref{s3} and the quantities
${\cal Q}_\xi$ obtained by the `counter-term subtraction method'.  As
noted in the Introduction, the quantities ${\cal Q}_\xi$ are
constructed from an `effective stress-energy tensor' which we will
denote by $\t_{ab}$ (to distinguish it from the stress energy-tensor
$\T_{ab}$ of physical matter fields). $\Q_\xi$ are defined by
\be\label{4.1}
{\cal Q}_{\xi} [C] = \lim_{{} \rightarrow \I} \oint_{\hat C} 
\t_{ab} {\hat \xi}^a d{\hat S}^b 
\ee
where $\hat{C}$ are $(d-2)$-spheres in $\M$ converging to the
cross-section $C$ of $\I$ and $\hat{\xi}^a$ is a vector field in $\M$
which tends to an infinitesimal asymptotic symmetry $\xi^a$ on
$\I$. In this work it is (implicitly) assumed that there is no
physical matter field in a neighborhood of infinity. Therefore, in
this section, we set $\T_{ab} =0$. Also, since the expressions for
${\hat \tau}_{ab}$ become rapidly complicated in higher dimensions, we
restrict ourselves to $d=4$ and $d=5$, although our qualitative
conclusions hold also in higher dimensions.

Let $(\M,\g_{ab})$ be asymptotically \ads in the sense of our
definition and let $(M,g_{ab}, \O)$ be a conformal completion
satisfying our conditions.  Denote by $\I_\O$ the $(d-1)$-dimensional
sub-manifolds of $M$ on which $\O$ is constant.  In a sufficiently
small but finite neighborhood of $\I$, each $\I_\O$ is a time-like
surface.  Denote by $\h_{ab}$ the intrinsic metric induced on $\I_\O$
by $\g_{ab}$, by $\hat{\G}_{ab}$ its Einstein tensor and by $\K_{ab}$
the extrinsic curvature.  (Following the standard sign conventions
used in general relativity \cite{wald}, we will set $\K_{ab} =
\hat{h}_a{}^m \hat{h}_b{}^n \hat{\D}_m \hat{\eta}_n$, where
$\hat{h}_{ab}$ is the intrinsic metric and $\hat{\eta}^a$ the unit
outward radial normal on $\I_\O$. This sign convention is opposite to
the one used in \cite{bk}.)  Then, motivated by considerations of
\cite{by}, an explicit expression of $\t_{ab}$ was obtained in
\cite{bk}:
\be\label{4.2}
\t_{ab} = \frac{1}{8\pi \Gn}~\left[ \frac{\l}{n-3}~\hat{{\cal G}}_{ab} 
-\frac{n-2}{\l}~\hat{h}_{ab} - \K_{ab} + \K \hat{h}_{ab} \right]
\ee
We will now relate $\t_{ab}$ to the field $\E_{ab}$ we used to define
our conserved charges. Note first that, using the relation between the
Riemann tensors of $\g_{ab}$ and $\h_{ab}$, we can express the
electric part $\hat{E}_{ab}$ of the Weyl tensor $\hat{C}_{ambn}$ of
$\g_{ab}$ as:
\be\label{4.3}
{\hat E}_{ab} := {\hat C}_{ambn}~{\hat \eta}^m {\hat \eta}^n 
= -{\hat {\cal R}}_{ac} + {\K}{\K}_{ac} - {\K}_{ab}
{\hat K}^b_c + \frac{2}{d-1}~\Lambda {\hat h}_{ab} 
\ee
Hence, it follows that
\ba\label{4.4}
8\pi \Gn ~\t_{ab} + \frac{\l}{d-3}~{\hat E}_{ab} 
&=&\frac{\l}{d-3} \left[ {\K} {\K}_{ab} - {\K}_a^{~c}~
{\K}_{bc} + \frac{1}{2} ( {\K}^{mn} {\K}_{mn} -
{\K}^2) {\h}_{ab} \right] \nonumber \\
&-&\frac{d-2}{2\l}~{\h}_{ab} - {\K}_{ab} + {\K}{\h}_{ab}
\ea

To take limit to $\I$, let us express the relevant `hatted' fields
(constructed from $\g_{ab}$) in terms of the corresponding `unhatted'
fields (constructed from $g_{ab}$). First, it is straightforward to
show that
\be\label{4.5}
{\K}_{ab} = \O^{-1}{K}_{ab} - \Omega^{-2} (\eta \cdot n) h_{ab}
\ee
where, as before, $n_a = \D_a \O$ and $\eta^a$ is the unit outward normal 
to $\I_\O$ with respect to $g_{ab}$.  Next, using (\ref{2.13}) and 
(\ref{2.10}), it is easy to show that the bold-faced fields in
\be\label{4.6}
{\bf K}_{ab} = \Omega^{-1} K_{ab}, \quad {\rm and} \quad {\bf f} =
\Omega^{-2} \left( l (\eta \cdot n) +1 \right)
\ee
admit smooth limits to $\I$. Using these facts, one can rewrite
(\ref{4.4}) as:
\begin{eqnarray}\label{4.7}
8\pi \Gn ~\t_{ab} + \frac{\l}{d-3}~{\hat E}_{ab} &=& \Omega^2~[
\frac{\l}{d-3}\left( {\bf K}{\bf K}_{ab} - {\bf K}_a^{~c} {\bf K}_{bc} +
\frac{1}{2} ({\bf K}^{mn} {\bf K}_{mn} - {\bf K}^2) h_{ab} \right)
\nonumber \\
&+& {\bf f} ({\bf K}_{ab} - {\bf K} h_{ab} ) -  \frac{d-2}{2\l}~{\bf f}^2
{\bf h}_{ac} ] \nonumber   \\
&=&   \Omega^2 \Delta_{ab},~~~ {\rm say}\, .
\end{eqnarray}
By inspection, $\Delta_{ab}$ admits a smooth limit to $\I$.  Finally, since 
$\hat{\xi}^a$ admits a smooth limit $\xi^a$ to $\I$ and $d\hat{S}^b = 
\O^{3-d} dS^b$, we can express the difference between the two sets of 
charges as:
\be\label{4.8} 
{\Q}_{\xi} [C] - Q_\xi [C] = \frac{1}{8\pi \Gn} \oint_c
\Omega^{d-5}~ \Delta_{ab} \xi^a ds^b \ee
It follows immediately that the two charges agree if $d=4$ but they do
not if $d=5$. To our knowledge, in four dimensions the equality had
been established only in the case of Kerr-\ads space-times and the
focus was on energy, i.e., on the case when the infinitesimal symmetry
$\xi^a$ corresponds to the time translation isometry of this
space-time. Eq (\ref{4.8}) establishes the result in \textit{all}
asymptotically \ads space-times and for \textit{all} asymptotic
symmetries.

Let us investigate the situation in $d=5$ further. Let $(\M, \g_{ab})$
be the \ads space-time itself. Choose a rest-frame, express $\g_{ab}$
in the adapted chart, set $\O= 1/r$ and consider the resulting
conformal completion as in Section \ref{s2.1}. Then, it is easy to
verify that
\be\label{4.9}
{\bf K}_{ab} = \frac{1}{\l}( 1 + \l^2 \Omega^2)^{1/2} \D_a t
\D_b t \quad {\rm and} \quad  {\bf f} = \Omega^{-2} \left[ 1 - (1+ \l^2
\Omega^2)^{1/2} \right] \, .\ee
It therefore follows that $\Delta_{ab} = -(3/8)\l^3 h_{ab}$. Since
$\hat{E}_{ab} =0$ on $\M$, as noted before, all $Q_\xi[C]$ vanish
identically. Hence, if we choose $\hat\xi = \partial/\partial t$ the 
time translation adapted to our initial choice of rest frame, we obtain
\be\label{4.10}
{\Q}_t [C]  =  \frac{\l}{8\pi G}~\frac{3}{8}~A_{(3)}
\ee
where, as before, $A_{(3)}$ is the area of the unit 3-sphere.  This is
the `ground state energy' of the `counter-term subtraction method'
\cite{bk}.  

In the string theory literature \cite{bk,new}, quantities ${\Q}_\xi$
(and their higher dimensional versions) are generally interpreted as
the gravitational analogs of the symmetry generators in the confor
field theory. It is important to examine if they are viable as
candidate Hamiltonians generating asymptotic symmetries on the
gravitational side. Now, as noted in Section \ref{s3.2} the fact that
${\Q}_t$ fails to vanish in pure \ads space-time implies that $\Q_\xi$
are not defined in a covariant fashion.  Indeed, if there is a
well-defined, non-zero ground state energy, the obvious question is:
what rest frame does it refer to?  There is no preferred rest-frame in
pure \ads space-time and covariance prevents us from saying that the
right side of (\ref{4.10}) is the ground state energy in
\textit{every} rest frame.  Finally, from the perspective of classical
general relativity, it would have been rather strange if there were a
ground state energy in five dimensions but not in four. If this energy
is to have physical significance, one should be able to measure it in
terms of `planetary' (i.e. test particle) motions and it is difficult
to imagine such a qualitative difference in four and five dimensions.

In general asymptotically \ads space-times, the remainder term
$\Delta_{ab}$ has a more complicated form. Although we do not have a
definitive result, it seems unlikely that the extrinsic curvature
terms will always conspire to cancel each other out. If they do not,
the difference between $Q_\xi$ and $\Q_\xi$ would not be just a
`constant shift'. Then, there would be \textit{two} independent sets
of conserved quantities in asymptotically \ads space-times. This seems
implausible.  And indeed, there is another problem with the definition
of $\Q_\xi$. As noted in Section \ref{3.2}, the trace-free property of
$\E_{ab}$ is essential for $Q_\xi$ to be conserved.  Unfortunately,
$\Delta_{ab}$ is not trace-free. Hence, even though there is no
physical matter fields near infinity, the charges $Q_\xi[C]$ fail to
be conserved; their values depend on the choice of the cross-section
$[C]$. This fact can go unnoticed if one focuses on a
\textit{specific} conformal factor and on the asymptotic symmetries
$\xi^a$ which are Killing fields \textit {of the resulting}
$h_{ab}$. However, this procedure introduces an additional structure
and thus breaks covariance. As emphasized in Sections \ref{s2.1} and
\ref{s3.1}, in general there is no preferred metric on $\I$, and we
must deal squarely with the full equivalence class $[h_{ab}]$ of
conformally flat metrics.

\section{Discussion}
\label{s5}

In its original version \cite{m}, the ADS/CFT conjecture suggests that
two quite distinct theories are equivalent: i) ${\cal N}= 4$
supersymmetric $SU(N)$ Yang-Mills theory in four dimensional Minkowski
space-time; and, ii) type-IIB string theory in ten space-time
dimensions subject to the boundary conditions that the space-time is
asymptotically a product of a (metric) 5-sphere and 5-dimensional \ads
space-time, where the radius of the 5-spheres equals the \ads radius
$\l$.  (The parameters of the two theories are related by $\l = (4\pi
g_{\rm s}N)^{1/4} \sqrt{\hbar\alpha^\prime}$ where $g_{\rm s}$ is the
string coupling constant and $T = 1/2\pi \alpha^\prime$ is the string
tension.) Although the boundary conditions used on the gravitational
side are not of direct physical interest, the conjecture is
fascinating from a mathematical physics perspective. In this respect
it has had remarkable successes, especially in the perturbative
regimes \cite{m}.  In particular, all low energy states of
supergravity (corresponding to the type IIB string theory in the
conjecture) can be mapped to states in the Yang-Mills Hilbert space.
Furthermore, in the large $N$ limit with $g_{\rm s} N$ constant,
several interactions also agree: in this limit, 3-point functions can
be calculated either in the supersymmetric Yang-Mills theory
\textit{or} in supergravity. Finally, it was possible to show that the
entropy of Schwarzschild-\ads black holes is proportional to their
horizon area in the limit $r_s \gg \l$ where $r_s$ is the
Schwarzschild radius.

Because of such surprising agreements, hopes have been expressed that
this duality may enable one \textit{in particular} to describe
classical gravity (with asymptotically \ads boundary conditions)
entirely in terms of the flat space, supersymmetric Yang-Mills theory.
To test these ideas, one needs an appropriate dictionary between the
two theories.  A natural strategy is to begin with simple, basic
observables which have `obvious' correspondence in the two theories.
Since the two theories share symmetry groups, observables generating
these symmetries are natural candidates.  One needs to first calculate
these observables \textit{separately within each theory} and then
analyze if it is consistent to map one set to the other.

In this note, we obtained conserved charges $Q_\xi$ corresponding to
asymptotic symmetries $\xi$ \textit{purely from the gravitational
perspective}.  The field equations and the boundary conditions
defining asymptotically \ads space-times naturally led us to
differential identities which, upon integration, yielded the
expression of $Q_\xi$.  In four space-time dimensions, our charges
$Q_\xi$ agree with the quantities $\Q_\xi$ obtained by the
`counter-term subtraction method' \cite{bk}.  However, in five (or higher)
dimensions they do not.  Furthermore, we were able to give a general
argument to show that the `ground state energies' obtained in that
method can not result from any covariant procedure.  Therefore, from
the perspective of classical general relativity, $\Q_\xi$ appear to be
unacceptable as Hamiltonians generating asymptotic symmetries. On the
other hand, it has been argued that the results of the `counter term
subtraction method' on the gravitational side agree with those
obtained from the supersymmetric Yang-Mills theory and are in fact
related to such fundamental features of that theory as the existence
of a Casimir energy. Let us suppose this is the case.%
\footnote{However, it is possible that these arguments overlooked some
subtlety in the gauge theory. In the calculation of the Casimir
energy of a field confined inside a \textit{physical} box, the full
\ads invariance is broken because the box defines a rest
frame. However, it is not obvious that there is such a preferred rest
frame in the supersymmetric Yang-Mills theory used in the ADS/CFT
conjecture.}
Then, one would have to conclude that there is a tension between
gravity and gauge theories: the `natural' dictionary would be
incorrect.  This is somewhat unsettling because the problem occurs for
the most basic of observables on both sides. Furthermore, these are
essentially the only `explicitly known' diffeomorphism invariant
observables in non-perturbative general relativity.

An outstanding issue then is whether a consistent dictionary does
exist. Since any two separable Hilbert spaces are isomorphic, a useful
correspondence between two theories has to achieve much more than than
an isomorphism between their Hilbert spaces: the isomorphism should
map a complete set of observables in one theory to a complete set of
corresponding observables in the other. Therefore, it would not
suffice to just compare the ground states and their energies in the
two theories. One must also account for other differences between
$Q_\xi$ and $\Q_\xi$. Indeed, the difference $\Delta_{ab}$ in the
integrands of the two expressions does not appear to be just a
constant shift but may vary from one space-time to another, especially
in the non-stationary context. More importantly, while all ten $Q_\xi$
are conserved in the absence of fluxes of physical matter fields
across $\I$, the ten $\Q_\xi$ are not. Thus, our results seem to
provide guidelines as well as constraints on possible dictionaries.

We will conclude by listing two open problems on the gravitational
side.  First, it would be desirable to probe the structure of
$\Delta_{ab}$ in detail. Are there asymptotic equations which we have
overlooked that force a cancellation between some of the terms? A
simplification in the form of $\Delta_{ab}$ may be necessary in the
construction of a consistent dictionary. Second, our analysis is based
directly on asymptotic field equations.  It should not be difficult to
examine the situation using a covariant phase space formulation
\cite{abr,wz}. It should also be possible to construct these
quantities starting with a manifestly finite action and performing the
Legendre transform; indeed it would be surprising if infinite
subtractions are \textit{essential} in purely classical treatments of
Hamiltonians. In the asymptotically Minkowskian context, this seems
to be possible. The asymptotically \ads case is under investigation.

\section*{acknowledgments}

We would like to thank Vijay Balasubramanian, Gary Horowitz, Kirill
Krasnov, Parthasarathi Majumdar, Robert Wald and especially Donald
Marolf for correspondence.  This work was supported in part by the NSF
grants PHY94-07194, PHY95-14240, INT97-22514 and by the Eberly
research funds of Penn State.

\end{document}